\documentclass[prb,twocolumn,showpacs,superscriptaddress,amsmath]{revtex4}
\usepackage{graphicx}
\usepackage{amssymb}
\usepackage{citesort}
\usepackage{color}
\begin{document}
\title{Bond patterns and charge order amplitude in $\frac{1}{4}$-filled
 charge-transfer solids}
\author{R.T. Clay}
\email{r.t.clay@msstate.edu}
\affiliation{Department of Physics and Astronomy and HPC$^2$ Center for 
Computational Sciences, Mississippi State University, Mississippi State MS 39762}
\author{A.B. Ward}
\affiliation{Department of Physics and Astronomy and HPC$^2$ Center for 
Computational Sciences, Mississippi State University, Mississippi State MS 39762}
\author{N. Gomes}
\affiliation{ Department of Physics, University of Arizona
Tucson, AZ 85721}
\author{S. Mazumdar}
\affiliation{ Department of Physics, University of Arizona
Tucson, AZ 85721}
\affiliation{ Department of Chemistry, University of Arizona
Tucson, AZ 85721}
\date{\today}
\begin{abstract}
Metal-insulator transition accompanied by charge-ordering has been
widely investigated in quasi-one-dimensional conductors, including in
particular organic charge-transfer solids. Among such materials the
$\frac{1}{4}$-filled band charge-transfer solids are of strong
interest, because of the commensurate nature of the charge-ordering in
these systems. The period-four charge-order pattern
$\cdots$1100$\cdots$ here is accompanied by two distinct bond
distortion patterns, giving rise to bond-charge-density waves (BCDW)
of types 1 and 2. Using quantum Monte Carlo methods, we determine the
phase diagram within the extended Hubbard Hamiltonian that gives both
types 1 and 2 BCDW in the thermodynamic limit.  We further investigate
the effect of electron-electron and electron-phonon interactions on
the amount of charge disproportionation. Our results show that between
these two bond patterns, one (BCDW2) in general coexists with a large
magnitude charge order, which is highly sensitive to electron-phonon
interactions, while the other (BCDW1) is characterized by weak charge
order. We discuss the relevance of our work to experiments on several
$\frac{1}{4}$-filled conductors, focusing in particular on the
materials (EDO-TTF)$_2$X and (DMEDO-TTF)$_2$X with large amplitude
charge-order.
\end{abstract}

\pacs{71.10.Fd, 71.45.Lr, 74.70.Kn}\maketitle

\section{Introduction}

 Molecular charge transfer solids (CTS) are widely studied
 because of their many complex electronic states.  Small structural
 changes can lead to very different electronic behaviors. These effects
 have been studied extensively in the quasi-one dimensional CTS, in
 particularly for the $\frac{3}{4}$-filled (with density $\rho=0.5$
 holes per molecule) materials (TMTSF)$_2$X and (TMTTF)$_2$X, which
 become superconducting under the application of pressure.

The ground state of a one dimensional (1D) system of electrons with
coupled lattice degrees of freedom is an insulating Peierls
state. Quite generally at $\rho=0.5$ the ground state is a bond-charge
density wave (BCDW) with coexisting charge order (CO) and bond
distortion and can be described by a Hamiltonian with
electron-electron (e-e) and electron-phonon (e-p) interactions
\cite{Ung94a,Clay03a,Mazumdar00a}.  Experimentally, the properties of
BCDWs in quasi-1D $\rho=0.5$ CTS are observed to vary widely. In
systems with type 1 BCDW, hereafter BCDW1, there occur {\it two}
distinct transitions, a high temperature ($\approx$100 K)
metal-insulator (MI) transition followed by a low temperature (T
$\leq$ 20 K) magnetic transition to a spin-gapped or antiferromagnetic
state that coexists with CO with weak amplitude (we define the
amplitude of the CO as the difference in charge densities between the
charge-rich and charge-poor molecular sites.)  The most well known
examples of BCDW1 systems are in the (TMTTF)$_2$X family.
Systems with type 2 BCDW, hereafter BCDW2, are less widely
known. There occurs a single MI transition in these systems which is
accompanied by both a charge gap and a spin gap. Experimentally
determined CO amplitudes in these cases are rather large
\cite{Clay12a}. One member of the BCDW2 family is (EDO-TTF)$_2$PF$_6$,
where the MI transition temperature is 280 K and the CO amplitude is
known to be approximately 0.9:0.1
\begin{figure}[tb]
\centerline{\resizebox{2.75in}{!}{\includegraphics{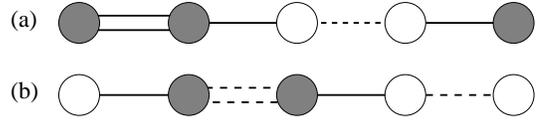}}}
\caption{Bond distortion patterns coexisting with $\cdots$1100$\cdots$
  CO in the $\frac{1}{4}$-filled band. Filled (unfilled) circles
  indicate molecules with charge density $0.5+\delta$ ($0.5-\delta$).
  (a) BCDW2 with bond distortion pattern Strong-Medium-Weak-Medium
  (SMWM). Here the double line indicates a stronger bond (S) than a
  single line (M), and solid lines are stronger than dashed lines (W).
  (b) BCDW1 with pattern Strong-Weak-Strong-Weak$^\prime$
  (SWSW$^\prime$). Here the single (S) bond is strongest, followed by
  double-dashed (W$\prime$) and single-dashed (W).}
\label{fig:cartoons}
\end{figure}
\cite{Ota02a,Drozdova04a,Aoyagi04a}. The large CO amplitude has led to
suggestions that interactions beyond e-e and e-p, such as molecular
bending \cite{Tsuchiizu08a} or electronic polarization effects
\cite{Iwano08a} are the driving forces behind the MI transition. Yet
another system that belongs to this class is (DMEDO-TTF)$_2$X, X =
ClO$_4$ and BF$_4$, where also there occurs a single MI transition
that opens both a charge and spin gap simultaneously. It has been
suggested that anion ordering drives the transition here
\cite{Kumeta16a}. The CO amplitude is currently unknown.  One goal of
our work is to show that both BCDW1 with small CO amplitude and BCDW2
with large CO amplitude can be understood within the same
one-dimensional fundamental theoretical model, albeit within different
parameter regions. The observed molecular bending \cite{Tsuchiizu08a}
as well as cation-anion interactions \cite{Kumeta16a} are consequences
and not the driving forces behind the {\it co-operative} transitions.

The bond distortion patterns corresponding to BCDW1 and BCDW2 are
shown in Fig.~\ref{fig:cartoons}.  In both the charge density follows
the pattern $\cdots$1100$\cdots$, where 1 (0) indicates a molecule
with charge density 0.5+$\delta$ (0.5-$\delta$). In BCDW2
(Fig.~\ref{fig:cartoons}(a)) the strongest bond is between the two
large charge densities (within the dimer), and the hopping integrals
follow the pattern strong-medium-weak-medium (SMWM).  In BCDW1
(Fig.~\ref{fig:cartoons}(b)), the pattern of hopping integrals in the
ground state is instead strong-weak-strong-weak$^\prime$
(SWSW$^\prime$) in Fig.~\ref{fig:cartoons}(b)), where the W$^\prime$
bond is slightly weaker than the W bond.

In this paper we perform a systematic numerical study of these two
states, with a goal of fully determining the phase diagram as well as
BCDW order parameters (amplitude of the CO and bond distortion) in the
thermodynamic limit.  We show that larger charge disproportionation
coexists with BCDW2, with magnitudes that are consistent with
experimental results. On the other hand, for BCDW1, we show that the
CO amplitude is significantly smaller.

The outline of the paper is as follows: in Section
\ref{section:results} we define the model and theoretical quantities,
followed by calculations for the phase diagram in
\ref{section:phasediagram}, and BCDW order parameters in
\ref{section:chg}. In Section \ref{section:discuss} we compare our
results with experimental studies of several materials.

\section{Results}
\label{section:results}

A well established minimal model for the 1D CTS
is the 1D Peierls-extended Hubbard  model,
\begin{eqnarray}
  H&=&-\sum_{i\sigma}[t-\alpha\Delta_i](c^\dagger_{i+1,\sigma}c_{i,\sigma} + H.c.)+\frac{1}{2}K_1\sum_i \Delta_i^2 \nonumber \\
  &+& g \sum_i \nu_i n_i + \frac{1}{2}K_2 \sum_i \nu_i^2 \nonumber \\
&+& U\sum_in_{i,\uparrow}n_{i,\downarrow} +V\sum_in_{i+1}n_i . \label{eq:ham}
\end{eqnarray}
In Eq.~\ref{eq:ham}, $c^\dagger_{i,\sigma}$ ($c_{i,\sigma}$) creates
(annihilates) an electron of spin $\sigma$ on site $i$,
$n_{i,\sigma}$=$c^\dagger_{i,\sigma}c_{i,\sigma}$, and
$n_i$=$n_{i,\uparrow}+n_{i,\downarrow}$. $\Delta_i$ is the deviation
of the bond between sites $i$ and $i+1$ from its equilibrium length and
$\alpha$ is the inter-site e-p coupling with spring
constant $K_1$.  Intra-molecular distortions on each molecule are
parameterized by the phonon coordinate $\nu_i$; $g$ is the intra-site
e-p coupling with $K_2$ its corresponding spring constant.  $U$ and
$V$ are the onsite and nearest-neighbor Coulomb interactions
respectively. We give energies in units of $t$.

At $\frac{1}{4}$-filling ($\rho=0.5$), charge- and bond-ordering at 2k$_{\rm
  F}$ (period four) or 4k$_{\rm F}$ (period two) dominate.  The occurrence
of 4k$_{\rm F}$ CO requires $V>V_c$, where the critical value\cite{Mila93a} $V_c=2$
in the limit $U\rightarrow\infty$ but is larger than $2$ for finite
$U$ (see Fig.~\ref{fig:phasediag}).
In applying Eq.~\ref{eq:ham} to the 1D CTS, it is also expected that
$V<\frac{U}{2}$, based on
comparison to $\rho=1$ 1D CTS \cite{Clay07a}.  Here we
restrict our analysis to regions of the phase diagram with
$V<V_c$ and $V<\frac{U}{2}$.  Throughout this region bond ordering is
in general a mixture of period four and period two distortions.  A general
form for $\Delta_j$ can  be written as \cite{Ung94a}
\begin{equation}
\Delta_j = \Delta_0[a_2\cos(2k_{\rm F}j-\phi_2)+a_4\cos(4k_{\rm F}j-\phi_4)],
\label{eq:r2r4}
\end{equation} 
where $\Delta_0$ is the overall amplitude of the bond distortion,
 $a_2$ and $a_4$ are the amplitude of 2k$_{\rm  F}$ 
and 4k$_{\rm F}$ components respectively, and $\phi_2$ and $\phi_4$
their phases.

Exact diagonalization solutions of Eq.~\ref{eq:ham} have found several
possible BCDW states \cite{Ung94a,Clay03a}.  In the region of phase
space we consider two different BCDW solutions are found, shown
schematically in Fig.~\ref{fig:cartoons}(a) and (b) and labeled BCDW2
and BCDW1 below.  Note that a second 2k$_{\rm{F}}$ charge pattern,
$\cdots$2000$\cdots$ is also possible, but only in the limit of very
weak e-e interactions\cite{Ung94a}; we will not consider it here.

\begin{figure}[tb]
  \resizebox{3.0in}{!}{\includegraphics{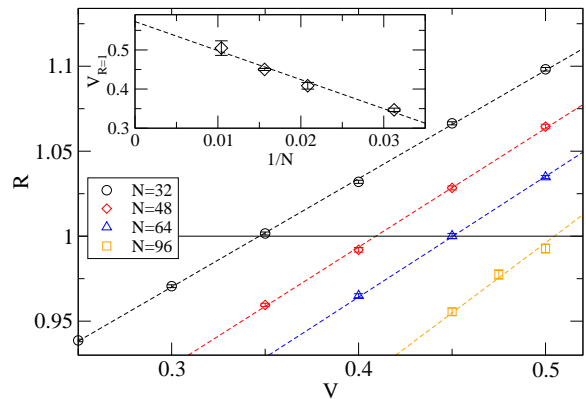}}
\caption{(color online)
  (a) $R=\chi_B(4k_{\rm F})/\chi_B(2k_{\rm F})$ as a function of $V$ with $U=6.25$.
  Circles, diamonds, triangles, and squares
  are for 32, 48, 64, and 96 site chains, respectively.
The inset shows the finite-size scaling of $V_{R=1}$, the $V$ for
  which $R=1$}.
\label{fig:fss}
\end{figure}

\subsection{Phase diagram}
\label{section:phasediagram}

While several previous works have demonstrated the presence of BCDW2
and BCDW1 in small-lattice exact diagonalization calculations, the
parameter regions of these two phases have not been mapped out in the
thermodynamic limit.  Here we determine the phase boundary between the
BCDW2 and BCDW1 in the limit of $0^+$ e-p coupling, i.e. the phase
boundaries that occur unconditionally for a given $U$ and $V$ in
the thermodynamic limit.

In Eq.~\ref{eq:r2r4} the phase angles for both BCDW states
are\cite{Ung94a} $\phi_2=\frac{\pi}{2}$ and $\phi_4=0$.  While BCDW2
is nearly a pure 2k$_{\rm{F}}$ bond distortion, BCDW1 requires a
significant 4k$_{\rm{F}}$ component.  The minimum $a_4$ in
Eq.~\ref{eq:r2r4} for the BCDW1 pattern occurs when the `S' and
`W$^\prime$' bonds are of equal strength.  From this one can derive
the condition that $a_4/a_2>\frac{1}{2}$ in the BCDW1 phase
\cite{Ung94a,note1}.  Further assuming the normalization $a_2+a_4=1$,
this implies $a_4>\frac{1}{3}$ for BCDW1.

The tendency to bond distortion at a wavevector $q$ is measured by the
bond susceptibility \cite{Hirsch84a}, $\chi_B(q)$, defined as
 \begin{equation}
\chi_B(q)=\frac{1}{N}\sum_{j,l}\int_0^\beta e^{iq(j-l)}\langle \tilde{B}_j(\tau)
\tilde{B}_l(0) \rangle d\tau.
\label{eq:suscep}
\end{equation}
In Eq.~\ref{eq:suscep}, $\tilde{B}_j(\tau)=e^{-\tau
  H}\tilde{B}_je^{\tau H}$ where $\tilde{B}_j=B_j-\langle B\rangle$
and $B_j=\frac{1}{2}\sum_\sigma
(c^\dagger_{j+1,\sigma}c_{j,\sigma}+H.c.)$.  $\beta$ is the inverse
temperature and $N$ the number of sites.  The BCDW2/BCDW1 phase
boundary corresponds to a specific ratio of 4k$_{\rm{F}}$ to
2k$_{\rm{F}}$ bond distortion and may therefore in the limit of $0^+$
e-p phonon coupling be determined by comparing $\chi_B(2k_{\rm F})$
and $\chi_B(4k_{\rm F})$.  The discrete Fourier transform of
$\Delta_j\Delta_l$ with respect to $(j-l)$ is $\Delta_0^2a^2_2N/4$ at
$q=\pi/2$ and $\Delta_0^2a^2_4N$ at $q=\pi$.  Therefore, in the limit
of $0^+$ e-p coupling $\chi_B(4k_{\rm F})/\chi_B(2k_{\rm
  F})=4a_4^2/a_2^2$ and the BCDW1 phase will occur when
$\chi_B(4k_{\rm F})/\chi_B(2k_{\rm F}) > 1$.  The bond distortion
changes smoothly between the two phases without any discontinuity in
the bond distortion or other observables.
\begin{figure}[tb]
\centerline{\resizebox{3.2in}{!}{\includegraphics{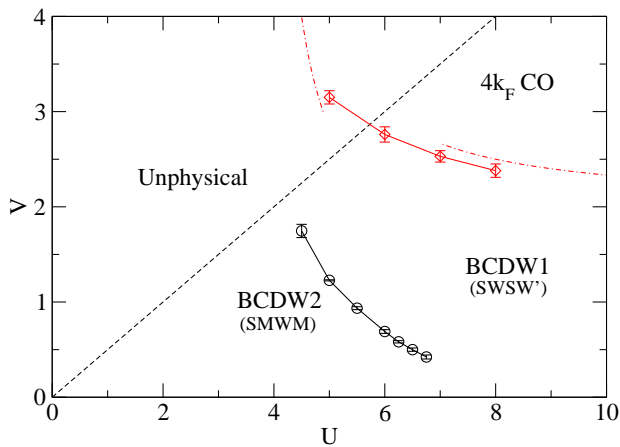}}}
\caption{(color online) Zero temperature phase diagram of Eq.~\ref{eq:ham} in the limit
  of $0^+$ e-p interactions at $\frac{1}{4}$ filling. Open points are
  the boundary between BCDW2 and BCDW1 regions. Diamonds and dot-dashed lines
mark the boundary to the
   4k$_{\rm{F}}$ CO region (see Ref.~\onlinecite{Clay03a}). The dashed
  line indicates the region of physical relevance for organic
  CTS, $V \alt \frac{U}{2}$ .}
\label{fig:phasediag}
\end{figure}

We use the Stochastic Series Expansion (SSE) quantum Monte Carlo
method with directed loop updates to calculate $\chi_B(q)$
\cite{Sandvik92a,Syljuasen02a}.  SSE is free from the Fermion sign
problem in 1D and provides exact (within statistical errors) results
at finite temperatures. We calculated the ratio $R=\chi_B(4k_{\rm
  F})/\chi_B(2k_{\rm F})$ for periodic systems of $N=$ 32, 48, 64, and
96 sites with an inverse temperature of $\beta=4N$, which is a low
enough temperature to give essentially ground state results.
$\chi_B(4k_{\rm{F}})$ increases with increasing $V$; for each system
size, the $V$ where $R=1$ was determined keeping $U$ fixed, as shown
in Fig.~\ref{fig:fss}.  We then performed a finite-size scaling using
a linear fit of the transition points to $1/N$; a typical fit is shown
in the inset of Fig.~\ref{fig:fss}.  Fig.~\ref{fig:phasediag} shows
the complete phase diagram in the ($U$,$V$) plane.  In
Fig.~\ref{fig:phasediag} we also include the boundary for the 4k$_F$
CO phase\cite{Mila93a} from Reference
\onlinecite{Clay03a}, which are determined from the condition that the
Luttinger Liquid exponent $K_{\rho}>\frac{1}{3}$, indicating dominant
4k$_{\rm{F}}$ charge fluctuations \cite{Voit95a}. The dot-dashed lines
in Fig.~\ref{fig:phasediag} are the result of second order
perturbation theory about the $U\rightarrow\infty$ and
$V\rightarrow\infty$ limits \cite{Mila93a,Lin00a}.
 In the rest of the paper we focus on 
the regions of the phase diagram occupied by BCDW1 and BCDW2. The phase 
boundary between 4k$_{\rm F}$ CO and BCDW1 has been discussed extensively in our 
previous work \cite{Clay03a,Clay07a}.

\subsection{Charge order amplitude}
\label{section:chg}

\begin{figure}[tb]
\centerline{\resizebox{3.0in}{!}{\includegraphics{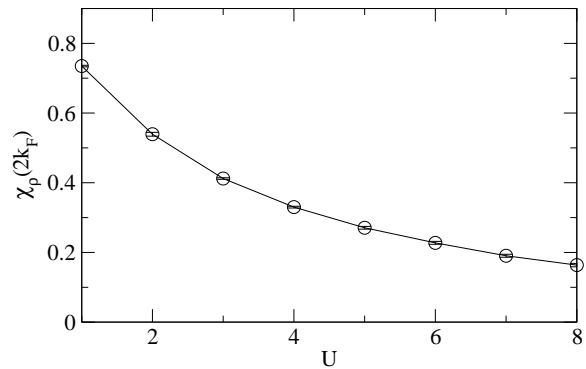}}}
\caption{2k$_{\rm{F}}$ charge susceptibility  as a
  function of $U$ for a 48 site chain with $V=U/4$, $\alpha=g=0$,
  and inverse temperature $\beta=192$.}
\label{fig:chg}
\end{figure}
\begin{figure}[tb]
\centerline{\resizebox{3.25in}{!}{\includegraphics{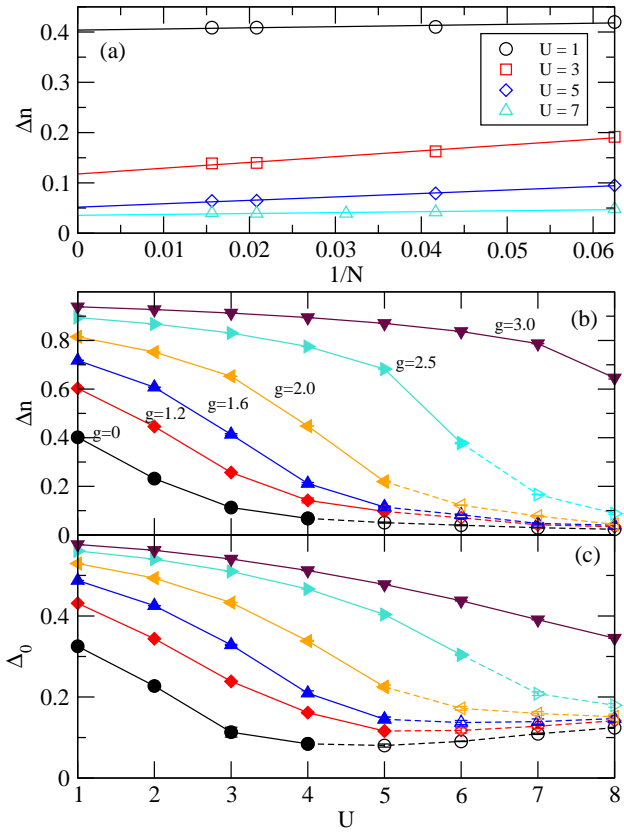}}}
\caption{(color online) Results of self-consistent MPS calculations (see text).
  For all panels $V=U/4$.
  (a) Finite size scaling of the charge order amplitude $\Delta n$
  versus inverse chain length with $\alpha=1.2$ and $g=0$. Lines are linear
  fits.
  (b) Finite-size scaled $\Delta n$ as a function of $U$ and $g$,
  with  $\alpha=1.2$.
  (c) The overall amplitude of the bond 
distortion (see Eq.~\ref{eq:r2r4}) for the parameters of (b). In both 
(b) and (c) the filled (open) points correspond to BCDW2 (BCDW1) and lines
are guides to the eye.}
\label{fig:mps}
\end{figure}
\begin{figure}[tb]
\centerline{\resizebox{3.25in}{!}{\includegraphics{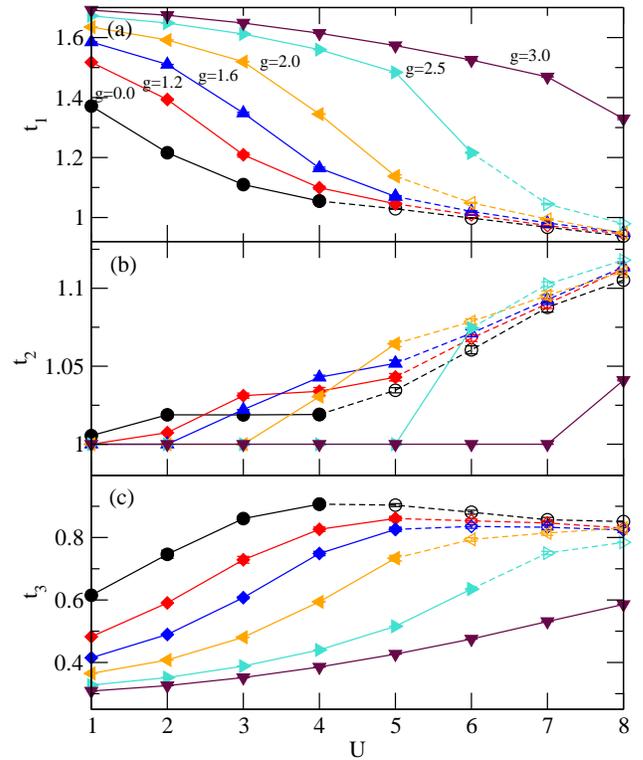}}}
\caption{(color online) Hopping integrals for the same parameters as
 Fig.~\ref{fig:mps}.}
\label{fig:mps-t}
\end{figure}

We define the amplitude of the CO as $\Delta$n = $\langle n_{\rm large}\rangle$
- $\langle n_{\rm{small}}\rangle$, where $n_{\rm large}$ and $n_{\rm{small}}$ are the charge densities on the
charge-rich and charge-poor molecules. $\Delta$n is of great experimental interest
and can be measured  optically \cite{Dressel12a} and by  NMR \cite{Zamborszky02a}. 
Theoretically, $\Delta$n is difficult to predict 
from Eq.~\ref{eq:ham}, as it depends on the precise values of the e-p
coupling constants $\alpha$ and $g$ which are difficult to estimate.

In the limit of $\alpha=g=0$, the 2k$_{\rm{F}}$ charge susceptibility
($\chi_\rho(q)$) is defined as in Eq.~\ref{eq:suscep} with
$\tilde{B}_j$ replaced by $n_j -\langle n\rangle$) decreases with
increasing $U$ \cite{Hirsch83a}, implying that $\Delta n$ is smaller
in BCDW1 compared to BCDW2. In Fig.~\ref{fig:chg} we show
$\chi_\rho(2k_{\rm{F}})$ as a function of $U$ calculated along the
line $V=U/4$ which crosses the BCDW2/BCDW1
boundary. Fig.~\ref{fig:chg} shows that differences in e-e correlation
alone can account for approximately a factor of four in the magnitude
of $\Delta n$ between the most weakly-correlated CTS salts compared to
those with strong e-e correlations, assuming equal e-p coupling
strengths.

To calculate $\Delta$n in Eq.~\ref{eq:ham} with e-p interactions, we
use a zero temperature variational quantum Monte Carlo using a
matrix-product state basis (MPS-QMC)
\cite{Sandvik08a,Clay12a}. Matrix-product states are extremely
efficient for representing the wavefunctions of interacting 1D
quantum systems. The MPS-QMC method variationally optimizes the MPS
matrices from random starting values using stochastic optimization
\cite{Sandvik08a}.
One advantage of MPS-QMC is that periodic systems can be easily treated.
Further details of the method are given in
Reference \onlinecite{Clay12a}. To handle the e-p degrees of
freedom self-consistently, $\Delta_i$ in Eq.~\ref{eq:ham} is taken
to be of the form of Eq.~\ref{eq:r2r4} with fixed $\phi_2$ and $\phi_4$.
Fixing the bond distortion to this form is reasonable provided
$U$ and $V$  are  restricted to the BCDW2/BCDW1 region of
the phase diagram---i.e. not too close to the 4k$_{\rm{F}}$ CO
region.
$\nu_i$ are taken with a constant magnitude $\nu$ and a
fixed pattern $\cdots --++ \cdots$ giving
$\cdots$1100$\cdots$ CO. Self-consistency equations for $\Delta_0$, $a_4$,
and $\nu$ are determined from \cite{Clay03a}
$$
\frac{\partial\langle H\rangle}{\partial\Delta_0}=0 \qquad \frac{\partial\langle H\rangle}{\partial a_4}=0
\qquad \frac{\partial\langle H\rangle}{\partial\nu}=0.
$$
For the results presented here, matrix dimensions $D$ of up to 32
were used (see Reference \onlinecite{Clay12a}). We used chain lengths
from 16 up to 64 sites and finite-size scaled the results using linear
extrapolation in $1/N$; Fig.~\ref{fig:mps}(a) shows typical
finite-size extrapolations for the case $\alpha=1.2$ and $g=0$.

The intra-site e-p interaction couples directly to the charge density
and affects $\Delta n$ strongly. We first choose a fixed $\alpha$ and
vary $g$ in Eq.~\ref{eq:ham}.  Figs.~\ref{fig:mps}(b) and (c)
summarize the results of these calculations.  For $g \alt 2$, $\Delta
n$ versus $U$ has a very similar functional shape as the 2k$_{\rm{F}}$
charge susceptibility in Fig.~\ref{fig:chg}, confirming that e-e
interactions strongly affect $\Delta n$.  The maximum $\Delta n$
for $g=0$ is $\approx 0.4$ at small $U$. As seen in
Figs.~\ref{fig:mps}(b) and (c), the bond pattern switches to BCDW1 at
$U \approx 5$, which is consistent with the phase diagram in
Fig.~\ref{fig:phasediag}.

As shown in Fig.~\ref{fig:mps}(b),  in the BCDW2 phase,
$\Delta n$ is strongly enhanced by $g$ up to nearly complete
charge transfers of $\Delta n \approx 0.9$. BCDW1 however is
characterized by small $\Delta n$ for all $g$, which for most parameters choices is
$\alt 0.1$.  While in general weaker e-e correlations coincide with
larger $\Delta n$, Fig.~\ref{fig:mps}  shows that as $g$
increases the phase boundary between BCDW2 and BCDW1 moves to larger
$U$ and $V$ (i.e. the BCDW2/BCDW1 phase boundary in
Fig.~\ref{fig:phasediag} moves towards the 4k$_{\rm{F}}$ CO phase with
increasing $g$).  Fig.~\ref{fig:mps} also shows that large enough $g$
suppresses the BCDW1 phase altogether. It is also possible that
large $g$ in combination with $U$ and $V$ near the 4k$_{\rm{F}}$ CO
phase results in $\cdots$1010$\cdots$ CO \cite{Clay03a}.

Importantly, at large $U$, the strength of the bond distortion behaves differently from
$\Delta n$.
While the amplitude of the $\cdots1100\cdots$ CO decreases continuously as the strength
of e-e interactions increases, Fig.~\ref{fig:mps}(c) shows that
for weaker e-p interactions,
the overall bond distortion strength $\Delta_0$ first reaches a
minimum at $U\approx 5$ and then  {\it increases} again for larger $U$.
The reason for this apparently counter-intuitive behavior
is that while the 2k$_{\rm{F}}$ bond distortion decreases with increasing $U$,
the 4k$_{\rm{F}}$ distortion increases with $U$ (and $V$), causing the
increase in $\Delta_0$.
\begin{figure}[tb]
\centerline{\resizebox{3.25in}{!}{\includegraphics{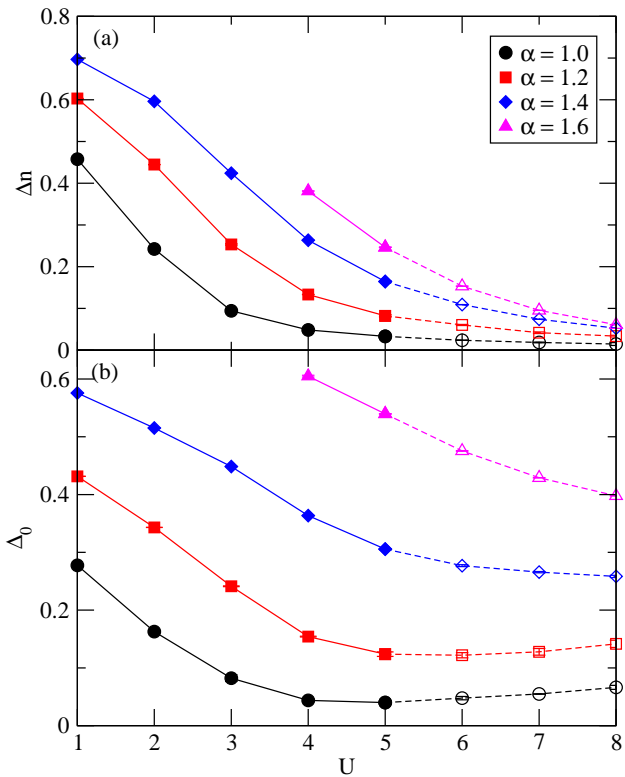}}}
\caption{(color online)   (a) Finite-size scaled $\Delta n$ as a function of $U$
  and $\alpha$  with $g=1.2$.
  (c) The overall amplitude of the bond distortion (see Fig.~\ref{fig:mps}(c)).}
\label{fig:mps2}
\end{figure}
\begin{figure}[tb]
\centerline{\resizebox{3.25in}{!}{\includegraphics{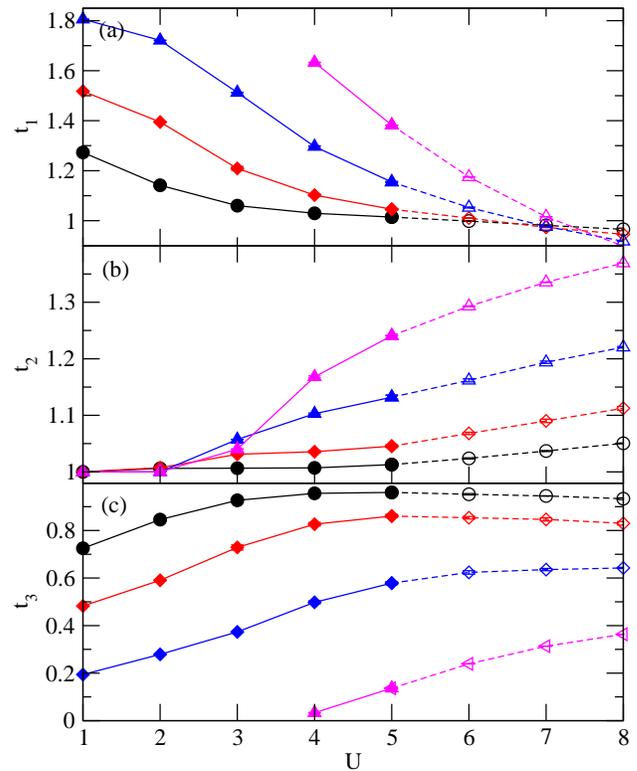}}}
\caption{(color online) Hopping integrals for the same parameters as
 Fig.~\ref{fig:mps2}.}
\label{fig:mps2-t}
\end{figure}

In the interest of comparing with experimental data, in
Fig.~\ref{fig:mps-t} we show the actual hopping
integrals. Corresponding to the charge order pattern
$\cdots1100\cdots$ we define the `1$-$1' bond as $t_1$, the `1$-$0'
and `0$-$1' bonds as $t_2$ and the `0$-0$0' bond as $t_3$,
respectively.  In the BCDW2 pattern SMWM, $t_1$ is the strong S bond,
$t_2$ the M bond, and $t_3$ the W bond.  In the BCDW1 pattern
SWSW$^\prime$, $t_1$ is the W$^\prime$ bond, $t_2$ the S bond, and
$t_3$ the W bond.  The decrease in $t_1$ in Fig.~\ref{fig:mps-t}(a)
and the simultaneous increase in $t_2$ in Fig.~\ref{fig:mps-t}(b) are
signatures of the crossover from BCDW2 to BCDW1 with increasing $U$.

In Fig.~\ref{fig:mps2} we show the result of varying the inter-site
e-p coupling $\alpha$.  Unlike $g$, $\alpha$ can only be varied
over a relatively small range. For finite systems a minimum value of
$\alpha$ is required for the lattice distortion to occur. For {\it too} large
$\alpha$ the linear e-p coupling in Eq.~\ref{eq:ham} leads to a
negative bond order for the weakest bonds indicating a failure of the
linear coupling assumption (this occurs for $\alpha=1.6$ and $U<4$ in
Fig.~\ref{fig:mps2}) \cite{Clay03a}.  
Fig.~\ref{fig:mps2} shows that varying $\alpha$
has a similar effect to varying $g$: stronger e-p coupling can enhance
$\Delta n$ strongly in the BCDW2 region, and at the same time moves
the system towards the BCDW2 phase.  Fig.~\ref{fig:mps2-t} further
shows the hopping integrals in this case. Increasing $\alpha$ can
strongly increase the amplitude of the bond distortion in BCDW1 (see
strong increase in Fig.~\ref{fig:mps2-t}(b)), however $\Delta n$
remains small.

Summarizing our data, in the BCDW2 region $\Delta n$ can have any
value up to $\approx 0.9$ depending on the e-e and e-p interaction
strengths.  However, regardless of the choice of e-e interactions and
e-p coupling strength, $\Delta n$ in the BCDW1 region is always
small---the maximum in all of our calculations was $\Delta n \approx
0.2$. More typically $\Delta n$ in the BCDW1 region is in the range
0.05 $\sim$ 0.1.

\section{Discussion}
\label{section:discuss}

\subsection{BCDW2}

Below we discuss two families of CTS whose low-temperature insulating
states show the BCDW2 bond pattern, (EDO-TTF)$_2$X (X = PF$_6$ and
AsF$_6$) and (DMEDO-TTF)$_2$X (X = ClO$_4$ and BF$_4$).  In each of
these materials the MI transition has been attributed to different
effects, such as e-p and coupled molecular bending \cite{Tsuchiizu08a}
or electrical potential bias \cite{Iwano08a} in (EDO-TTF)$_2$X, and
anion ordering in (DMEDO-TTF)$_2$X \cite{Kumeta16a}.  We argue that
these are instead {\it cooperative} effects, which are particularly
obvious due to the large magnitude of $\Delta n$ and bond distortion
found in BCDW2.  We expect that other 1D CTS with the BCDW2 distortion
will show similarly strong effects at the MI transition.  While these
secondary effects will enhance the amplitude of the BCDW, they are not
the principal driver of the transition, which is instead the
underlying tendency to distortion of the quasi-1D electron system. The
features common to BCDW2 are clearly seen by comparing (EDO-TTF)$_2$X
and (DMEDO-TTF)$_2$X, which show the same molecular stack distortion
but quite different secondary effects.

{\it{(EDO-TTF)$_2$X:}} In (EDO-TTF)$_2$X the MI transition is first
order and occurs at 280 K and 268 K for X=PF$_6$ and AsF$_6$,
respectively \cite{Ota02a}. A third salt, X=ClO$_4$, has an even
higher transition temperature, greater than 337 K \cite{Ota03a}.  In
(EDO-TTF)$_2$PF$_6$ the experimentally estimated CO amplitude is
rather large, with estimates of $\Delta n$ from optical measurements
of 0.92 ($T$=6 K) \cite{Drozdova04a} or from X-ray measurements of 0.6
($T=$260 K) \cite{Aoyagi04a}. Above the transition, the molecular
overlaps along the EDO-TTF stacks are nearly uniform with only a
slight dimerization \cite{Ota02a}. Below the transition the overlap
integrals follow the pattern SMWM \cite{Ota02a}.

Several observations indicate that intra-site e-p interactions are
strongly involved in the MI transition. At the transition, the EDO-TTF
molecules bend significantly \cite{Ota02a}, with the dihedral angles
changing by more than 5$^\circ$.  The position of the anions also
shift, with a periodic modulation that matches that of the EDO-TTF
stacks \cite{Ota02a}.  Optical studies of (EDO-TTF)$_2$X have
suggested that the observed high sensitivity to photoexcitation is
likely due to strong electron-lattice coupling \cite{Chollet05a}.

{\it{(DMEDO-TTF)$_2$X:}} Here the MI transition is at 190 K and 210 K
for X=ClO$_4$ and BF$_4$, respectively
\cite{Fabre95a,Kumeta16a}. Above the MI transition the organic
molecules are stacked uniformly, and like (EDO-TTF)$_2$X the low
temperature overlap integrals are in the SMWM pattern
\cite{Kumeta16a}.  Simultaneously with the stack distortion, the anion
positions shift, moving closer (further) towards molecules with large
(smaller) hole density.  The authors of Reference
\onlinecite{Kumeta16a} ascribe the MI transition to anion ordering, as
in (TMTSF)$_2$ClO$_4$.  Note, however, that in contrast to
(TMTSF)$_2$ClO$_4$ there is no simple rotational ordering of the
ClO$_4$ anion in (DMEDO-TTF)$_2$ClO$_4$.  Rather the Cl atom of the
ClO$_4$ group moves towards and away from charge-rich and charge-poor
molecules, which is a simple electrostatic effect.  While $\Delta n$
estimated from carbon-carbon bond lengths appears to be small (this
method of estimating CO amplitude however has large errors)
\cite{Kumeta16a}, we predict that optical measurements will find large
$\Delta n$ in this material.

To obtain the large $\Delta n$ found in (EDO-TTF)$_2$X, our results of
Section \ref{section:results} show that large intra-site e-p coupling
(and moderate or small e-e correlations) are required. The strong
coupling to molecular bending in (EDO-TTF)$_2$X shows that
intra-molecular modes are coupled strongly in this case. Similarly,
the large $\Delta n$ will lead to large potential energy differences
\cite{Iwano08a}.  In both (EDO-TTF)$_2$X and (DMEDO-TTF)$_2$X,
electrostatic effects will shift the position of the anions.
\medskip

\subsection{BCDW1}

As we have considered the thermodynamics of materials with the BCDW1
distortion in previous works \cite{Clay03a,Clay07a}, we will not
discuss them in detail here.  The BCDW1 state can be visualized as a
second dimerization of a dimer lattice. In this case two thermodynamic
transitions are expected, with the intermediate temperature state
having either dimerization or 4k$_{\rm F}$ CO \cite{Clay03a,Clay07a}.
What the present calculations show is that in the ground state, the
expected CO amplitude in BCDW1 is quite small and may be difficult to
detect experimentally. This should also be taken into consideration in
searches for CO in two dimensional CTS \cite{Sedlmeier12a}. To detect
the presence of BCDW1, it may be easier to focus on the pattern of
bond distortion rather than the amount of CO.

Empirically, CTS showing the BCDW1 distortion are more likely to show
superconductivity (SC) under pressure \cite{Clay12a}. We have
suggested that unconventional SC in $\rho=0.5$ materials arises from
the delocalization of the (`1$-$1') singlet pairs formed in the
insulating Paired Electron Crystal (PEC), which has the same CO
pattern $\cdots1100\cdots$ as BCDW1 and BCDW2
\cite{Mazumdar14a,Gomes16a,DeSilva16a}. Within such a model, the
effective mass of the singlet is smaller in the BCDW1 because of the
weaker binding, and hence the mobility of inter-dimer pairs here would
be expected to be large, allowing a transition to a paired liquid
state under the application of pressure \cite{Gomes16a}. On the other
hand, the intra-dimer pairs with large $\Delta n$ found in BCDW2
materials would tend to remain in an insulating state.

\section{Acknowledgments}

This work was supported by the Department of Energy grant
DE-FG02-06ER46315.  Part of the calculations were performed using
resources of the National Energy Research Scientific Computing Center
(NERSC), which is supported by the Office of Science of the
U.S. Department of Energy under Contract No. DE-AC02-05CH11231.


\begin{thebibliography}{29}
\bibitem[{not()}]{note1}
\bibinfo{note}{In References \onlinecite{Ung94a} and \onlinecite{Clay12a} the
  bond distortion pattern is written in terms of $u_j$, the displacement of the
  $j$th site from equilibrium, with 2k$_{\rm F}$ and 4k$_{\rm F}$ weights $r_2$
  and $r_4$. In the present notation $\Delta_j=u_{j+1}-u_j$. The ratio of
  4k$_{\rm F}$ to 2k$_{\rm F}$ components required for the BCDW1 phase was
  quoted incorrectly in these references--rather than 0.41, $r_4/r_2 >
  \frac{\sqrt{2}}{4+\sqrt{2}}=0.26$ for the BCDW1 pattern.}
\end{thebibliography}
\end{document}